 \documentclass[cameraready]{Interspeech}

\usepackage{cite}  
\usepackage{booktabs,tabularx,multirow,ragged2e,siunitx}
\usepackage{graphicx}
\usepackage{amsmath}
\usepackage{amssymb}  
\usepackage{tikz}
\usepackage{capt-of}
\usepackage{placeins}   
\usepackage{afterpage}       
\usepackage{xurl}
\usepackage{microtype}
\usepackage{threeparttable}
\usepackage{tablefootnote}
\usepackage{etoolbox}
\makeatletter

\renewenvironment{thebibliography}[1]{%
  \begin{oldthebibliography}{#1}%
  \setlength{\itemsep}{0pt}
  \setlength{\parsep}{0pt}
  \setlength{\parskip}{0pt}%
  \setlength{\topsep}{0pt}
}{%
  \end{oldthebibliography}%
}
\makeatother

\AtBeginDocument{
  \setlength{\abovedisplayskip}{2pt plus 1pt minus 1pt}
  \setlength{\belowdisplayskip}{2pt plus 1pt minus 1pt}
  \setlength{\abovedisplayshortskip}{2pt plus 1pt minus 1pt}
  \setlength{\belowdisplayshortskip}{2pt plus 1pt minus 1pt}
}

\newcolumntype{Y}{>{\RaggedRight\arraybackslash}X}
\title{UniWhisper: Efficient Continual Multi-task Training\protect\\
for Robust Universal Audio Representation}

\author[affiliation={1}, orcid=0009-0009-0389-8307, equalcontribution, correspondingauthor]{Yuxuan}{Chen}
\author[affiliation={2}, equalcontribution]{Peize}{He}
\author[affiliation={3}, equalcontribution]{Haoyuan}{Yu}
\author[affiliation={4}]{Junzi}{Zhang}

\address{
    $^1$ Jilin University \\
    $^2$ University of Electronic Science and Technology of China \\
    $^3$ Hunan University
    $^4$ Shandong University
}

\email{yxchen5522@mails.jlu.edu.cn, 2023300904027@std.uestc.edu.cn, y15352176976@hnu.edu.cn, 202300800615@mail.sdu.edu.cn}


\usepackage{comment}

\begin{document}

\maketitle

\begin{abstract}

A universal audio representation should capture fine-grained speech cues and high-level semantics for environmental sounds and music in a single encoder. However, prior encoders often excel in one domain but degrade in others. We propose \textbf{UniWhisper}, an efficient continual multi-task training framework that casts heterogeneous audio tasks into a unified instruction and answer format. This enables standard next-token training without task-specific heads and losses. We assess the encoder using shallow MLP probes and k-nearest neighbors (kNN) on 20 tasks spanning speech, environmental sound, and music with the entire framework trained on only 38k hours of public audio. UniWhisper reaches normalized weighted averages of 0.81 with MLP probes and 0.61 with kNN, compared to 0.64 and 0.46 for Whisper, while retaining strong speech performance.

\end{abstract}

\section{Introduction}

\afterpage{%
  \begingroup
  \setlength{\dblfloatsep}{2pt plus 1pt minus 1pt}    
  \setlength{\dbltextfloatsep}{2pt plus 1pt minus 1pt} 
  \setlength{\abovecaptionskip}{8pt plus 2pt minus 2pt} 
  \setlength{\belowcaptionskip}{0pt}

  \begin{figure*}[!t]
   \centering
   \includegraphics[width=\textwidth]{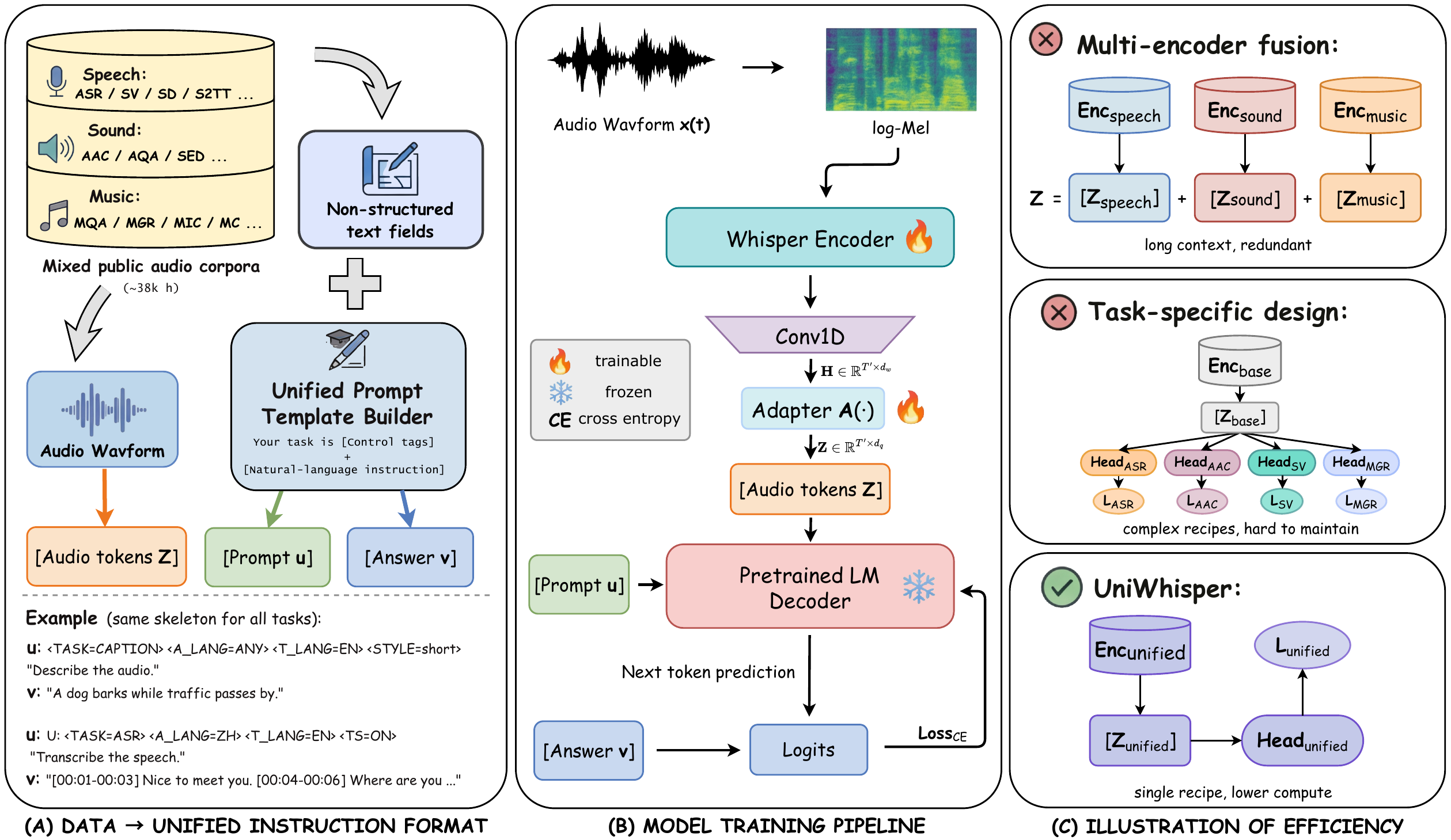}
   \captionsetup{font={stretch=0.8}}
   \caption{\textbf{Overview of the UniWhisper continual multi-task training framework.} \textbf{(A)} Converting heterogeneous datasets into a unified instruction and answer format. \textbf{(B)} Training with a single audio encoder and next-token prediction on answer tokens. \textbf{(C)} Comparison with common alternatives, highlighting reduced audio token redundancy and a unified supervision interface.}

   \label{fig:framework}
   \vspace{-3mm}
  \end{figure*}
  \endgroup
}
A universal audio representation aims to support speech, environmental sounds, and music with a single encoder. Large-scale pretraining has produced strong backbones for individual domains. Wav2vec 2.0~\cite{wav2vec2}, HuBERT~\cite{hubert}, WavLM~\cite{wavlm}, and Whisper~\cite{whisper} improve robustness and performance on speech-related tasks. BEATs~\cite{beats} and CLAP~\cite{clap} perform strongly on audio event recognition and audio-text semantic matching. However, this training progress also highlights a persistent domain imbalance. Speech-focused encoders often lack semantic coverage for complex non-speech scenes. In contrast, general audio models often fail to preserve the fine-grained temporal cues required by speech.

This split becomes especially costly in large audio language models (LALMs). A common recipe aligns a pretrained audio encoder with a large language model using paired audio and text data. When the encoder is speech-centric, broader audio coverage is often achieved through continual training on large and diverse datasets, which increases training cost and data requirements~\cite{af2,qwenaudio,qwen2audio,qwen25omni,stepaudio2}. Dual-encoder systems such as SALMONN~\cite{salmonn} and Kimi-Audio~\cite{kimiaudio} improve domain coverage by fusing encoders from different domains. However, they require additional coordination across temporal resolutions and representation spaces, and they often need extra alignment data. More importantly, concatenating features from multiple encoders increases the number of audio tokens and consumes the limited context window of the language model, while broader coverage often comes with longer token sequences.

\FloatBarrier 

\begin{figure}[!t]
  \centering
  \includegraphics[width=\linewidth]{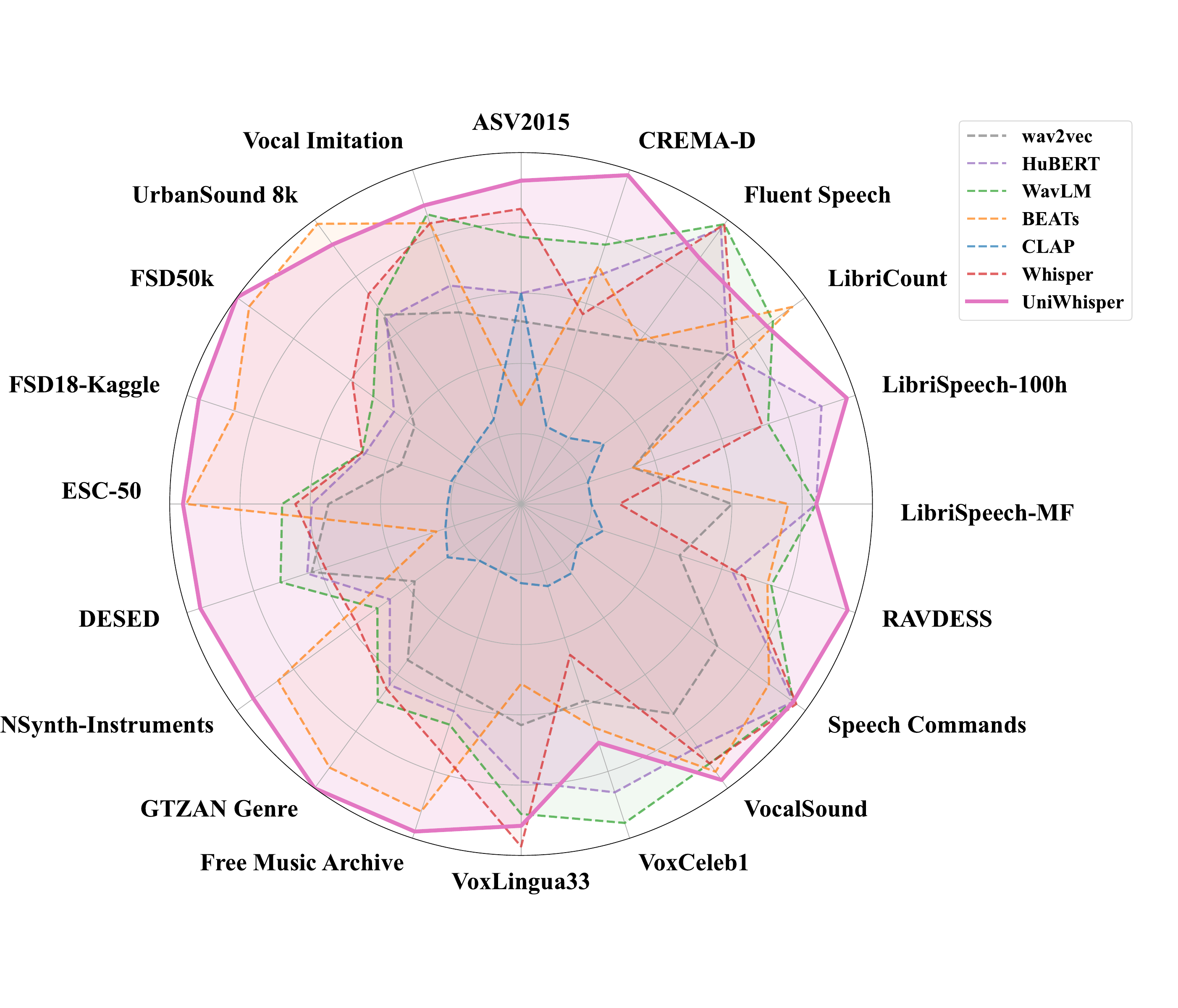}
  \vspace{-3.0mm}
  \captionsetup{font={stretch=0.8}}
  \caption{Normalized per-task performance of UniWhisper on our 20-task extended HEAR~\cite{hear2021} spanning speech, environmental sound, and music. Full results are reported in Table~\ref{tab:full_results}.}
  \vspace{-4.0mm}
  \label{fig:fig1}
\end{figure}

To avoid architectural redundancy, we adopt a single-encoder design and focus on unifying supervision. Whisper already learns rich acoustic perception from large-scale weakly supervised transcription, but transcription dominated training biases the representation toward speech. We propose prompt guided continual multi-task training, where diverse objectives are expressed with a shared instruction and answer format. This expands coverage across speech, environmental sound, and music without task specific heads or multi encoder feature concatenation. Since the audio prefix always comes from one encoder stream, token redundancy is removed at the source.

Using this framework, we train \textbf{UniWhisper}, a unified encoder backbone for multi-domain audio understanding that strengthens non-speech semantics while preserving speech capability such as ASR. We also identify an efficiency bottleneck in the original Whisper decoder under instruction-style alignment. In our setting, the decoder converges slowly and requires substantially more updates to reach competitive performance. We replace it with a compact pretrained language model that serves as the semantic interface during instruction-style training. The compact decoder provides strong language priors that better match instruction-following targets and can accelerate convergence. As a result, UniWhisper can be adapted with a substantially smaller training corpus than recent LALMs. We train on 38k hours of public audio, while Qwen2-Audio~\cite{qwen2audio} reports 520k hours for pre-training. The pipeline is simple because all tasks share the same templates and data format, so we can mix multiple open-source datasets with a single recipe. We assess representations using shallow MLP probing and non-parametric kNN evaluation, with ablations on the decoder and backbone. Experiments show that UniWhisper is competitive across 20 tasks spanning speech, environmental sound, and music. Under our evaluation protocol, it outperforms Whisper, HuBERT, BEATs, WavLM and CLAP on average and shows no clear catastrophic forgetting. Code and pretrained weights will be released.

\section{Method}

\subsection{Backbone and initialization}

Whisper uses an encoder-decoder transformer model. Given audio $x$, we extract log-Mel features and encode them into acoustic representations. The decoder predicts the next text token conditioned on the audio representation and previous tokens, trained with a standard cross-entropy loss.

UniWhisper retains the standard next-token prediction objective, but revises the decoder to better fit instruction-style supervision. The encoder is initialized from Whisper Large-v3~\cite{whisper}, while the autoregressive decoder is a pretrained language model Qwen3-0.6B~\cite{qwen3}. A lightweight adapter maps the encoder hidden states to the decoder hidden dimension.

Let $\mathbf{H}\in\mathbb{R}^{T' \times d_w}$ denote the Whisper encoder outputs and let $d_q$ be the hidden size of the decoder. The adapter produces
\begin{equation}
\mathbf{Z} = A(\mathbf{H}) \in \mathbb{R}^{T' \times d_q},
\end{equation}
where $A(\cdot)$ is a small MLP that performs a learned projection from $d_w$ to $d_q$. Using a pretrained LM as the decoder provides strong language priors that match instruction-following targets and can ease alignment between audio representations and text semantics. The pretrained LM decoder keeps frozen throughout training while the encoder and projection modules updating.

\subsection{Unified instruction style multitask training}

We express diverse audio tasks in a unified instruction and answer format, including ASR, speech translation, audio captioning, keyword or attribute prediction, audio question answering, and audio-text matching. As illustrated in Fig.~\ref{fig:framework}, each example is represented by a prompt that combines control tags, such as task type, audio language, text language, optional timestamps, and output constraints, with a natural language instruction. Supervision is provided as a text answer. Different tasks therefore differ only in the prompt content and the target answer, while sharing the same training and decoding interface.

Given the audio prefix $\mathbf{Z}$, prompt tokens $\mathbf{u}$, and answer tokens $\mathbf{v}=(v_1,\dots,v_{|\mathbf{v}|})$, the decoder models
\begin{equation}
p_\theta(v_t \mid v_{<t}, \mathbf{u}, \mathbf{Z}),
\end{equation}
and we optimize next token cross entropy on the answer tokens
\begin{equation}
\mathcal{L}_{\text{CE}} = -\sum\nolimits_{t=1}^{|\mathbf{v}|}\log p_\theta(v_t \mid v_{<t}, \mathbf{u}, \mathbf{Z}).
\end{equation}

Since UniWhisper uses a single audio token stream $\mathbf{Z}$ as the prefix, it avoids multi-encoder feature concatenation and the resulting audio token redundancy. Multi domain capability is obtained primarily through unified templates and mixed task training data under continual multi-task training.


\section{Experimental Setup}

\subsection{Datasets}

\textbf{Training:} we train UniWhisper on a mixture of open-source audio datasets summarized in Table~\ref{tab:datasets}.
We preserve the original supervision signals while converting all datasets into a unified instruction–answer format.
To prevent train–eval leakage, we apply identifier-based filtering and content-level deduplication using hashing and acoustic fingerprinting.

\begin{table}[t]
\centering

{\footnotesize 
\setlength{\tabcolsep}{3.2pt} 
\setlength{\aboverulesep}{0pt}
\setlength{\belowrulesep}{0pt}
\setlength{\abovetopsep}{1pt}
\setlength{\belowbottomsep}{1pt}
\setlength{\cmidrulesep}{0pt}

\captionsetup{font={stretch=0.8}}
\caption{Public datasets used for continual multi-task training, grouped by domain. Duration denotes the total audio hours used after filtering and deduplication.}
\vspace{-3.0mm}     
\label{tab:datasets}

\begin{tabularx}{\linewidth}{@{}l@{\hspace{4pt}}X@{\hspace{4pt}}l r@{}}
\toprule
Type & Name & Task & Duration \\
\midrule
\multirow{4}{*}{General}
& AudioCaps~\cite{audiocaps} & Captioning & 142.5~h \\
& AudioSet~\cite{audioset} & Tagging & 5.8k~h \\
& LAION-Audio~\cite{laionaudio630k} & Captioning & 4.3k~h \\
& WavCaps~\cite{wavcaps} & Captioning & 7.6k~h \\
\midrule
\multirow{4}{*}{Speech}
& AISHELL-1~\cite{aishell1} & ASR \& SID & 178~h \\
& GigaSpeech~\cite{gigaspeech} & ASR & 10k~h \\
& Libri-adhoc40~\cite{libriadhoc40} & ASR & 4.5k~h \\
& LibriSpeech~\cite{librispeech} & ASR \& SV & 860~h \\
\midrule
\multirow{6}{*}{Sound}
& Clotho~\cite{clotho} & Captioning & 43.6~h \\
& Seeing Sound~\cite{seeingsound} & SED & 0.2~h \\
& SONYC-UST~\cite{sonycustv2} & Tagging & 51.5~h \\
& TAU-ASC2020~\cite{tauasc2020} & ASC & 64~h \\
& URBAN-SED~\cite{urbansed} & SED & 27.8~h \\
& VGGSound~\cite{vggsound} & ASC & 553.3~h \\
\midrule
\multirow{9}{*}{Music}
& GuitarSet~\cite{guitarset} & Guitar Trans & 3~h \\
& MAESTRO~\cite{maestro} & Piano Trans & 200~h \\
& MedleyDB~\cite{medleydb} & AMT & 7.3~h \\
& MTG-Jamendo~\cite{mtgjam} & Tagging & 3.8k~h \\
& MusicCaps~\cite{musiccaps} & Captioning & 15.3~h \\
& MusicNet~\cite{musicnet} & AMT & 34~h \\
& Slakh2100~\cite{slakh2100} & AMT & 145~h \\
& SongDescriber~\cite{songdesc} & Captioning & 23~h \\
& YT8M-MTC~\cite{yt8mmtc} & Captioning & 11.7~h \\
\bottomrule
\end{tabularx}

\vspace{-4.0mm}
}
\end{table}

\noindent\textbf{Evaluation:} as shown in table \ref{tab:Evaluation_Protocol}, we build our evaluation set on HEAREval~\cite{hear2021}and  evaluate on 20 tasks spanning speech, environmental sound, and music.
Since HEAR provides limited coverage of human voice processing~\cite{xares}, we add speech-oriented tasks. More details refer to HEAR~\cite{hear2021} and X-ARES~\cite{xares}.

\subsection{Training details}

\textbf{Audio preprocessing:} We resample all audio to 16 kHz and extract 128-bin log-Mel features using a 25 ms window and a 10 ms hop. We use 30 s clips and pad shorter clips with zeros. To reduce the audio sequence length, we apply an additional temporal strided convolution with stride 2. Together with Whisper encoder subsampling, each output frame represents approximately 40 ms of the input waveform. We propagate padding masks through the encoder so that padded frames do not contribute to attention.

\noindent\textbf{Optimization:} We minimize next-token cross-entropy loss only on answer tokens while masking prompt tokens. We use 8-bit AdamW with cosine decay (LR $2\times10^{-5}$, weight decay $0.01$), a 1{,}500-step warm-up to align the audio and text representation spaces, and train for 30{,}000 update steps in total. Training uses bf16 and DDP under 24 wall-clock hours on 8 A800 GPUs with batch size 32 per GPU (global batch size 256). Unless otherwise specified, we update only the Whisper encoder and the projection adapter, keeping the pretrained LM decoder frozen.

\subsection{Evaluation protocols}

We evaluate encoder representations with two protocols: supervised probing with a shallow MLP and non-parametric kNN. Tasks are divided into clip tasks that use a single embedding per example and frame tasks that use a sequence of embeddings. We also include an ASR task\footnote {Inspired by X-ARES~\cite{xares}, we use Qwen2.5 0.5B as the text decoder for ASR, and keep it frozen during training.} on LibriSpeech-100h subset to check whether continual multi-task training preserves the speech recognition ability.
\textbf{Embedding:} We use frame embeddings from the final encoder layer. For clip tasks, we mean pool over time to obtain one clip embedding. For frame tasks, we keep the full frame sequence and align frame labels by padding when needed.
\textbf{MLP:} We freeze the encoder and train a shallow MLP on top of the clip embedding or the frame sequence for each task.
\textbf{kNN:} We perform classification or retrieval directly in the pooled embedding space without training a classifier.
\textbf{Metrics:} We map each metric to $[0,1]$ via min--max normalization using best/worst attainable values (Acc/mAP/Seg-F1/Recall@1 use $1/0$);
for ASR we use $\mathrm{iWER}=\max(1-\mathrm{WER},0)$.
We compute $S=\frac{\sum_i n_i\hat{M}_i}{\sum_i n_i}$ and report it for MLP and kNN in table~\ref{tab:Evaluation_Protocol}.


\begin{table}[t]
\centering
{\footnotesize 
\setlength{\tabcolsep}{4.6pt}   
\renewcommand{\arraystretch}{0.9}
\setlength{\aboverulesep}{0pt}
\setlength{\belowrulesep}{0pt}
\setlength{\abovetopsep}{2pt}
\setlength{\belowbottomsep}{2pt}
\setlength{\cmidrulesep}{0.5pt}

\captionsetup{font={stretch=0.8}}
\caption{Evaluation tasks, metrics, and weights used to compute normalized weighted averages. Weights are proportional to the number of test examples.}
\vspace{-3.0mm}
\label{tab:Evaluation_Protocol}

\begin{tabular*}{\linewidth}{@{\extracolsep{\fill}} ll c rr @{}}
\toprule
\multirow{2}{*}{Type} & \multirow{2}{*}{Task} & \multirow{2}{*}{Metric} & \multicolumn{2}{c}{Weight} \\[-0.35ex]
\cmidrule(lr){4-5}
 &  &  & {\scriptsize MLP} & {\scriptsize kNN} \\[-0.55ex]
\midrule
\addlinespace[0.1pt]

\multirow{11}{*}{Speech}
& ASV2015                 & Acc   & 2000  & 2000 \\
& CREMA-D                 & Acc   & 1116  & 1116 \\
& Fluent Speech Commands  & Acc   & 2000  & 2000 \\
& LibriCount              & Acc   & 1144  & 1144 \\
& LibriSpeech-100h        & iWER  & 10000 &  -   \\
& LibriSpeech-MF          & Acc   & 2620  & 2620 \\
& RAVDESS                 & Acc   & 360   & 360  \\
& Speech Commands V1      & Acc   & 2000  & 2000 \\
& VocalSound              & Acc   & 2000  & 2000 \\
& VoxCeleb1               & Acc   & 2000  & 2000 \\
& VoxLingua33             & Acc   & 1609  & 1609 \\

\midrule
\addlinespace[0.6ex] 

\multirow{3}{*}{Music}
& Free Music Archive Small & Acc  & 800   & 800  \\
& GTZAN Genre              & Acc  & 100   & 100  \\
& NSynth-Instruments       & Acc  & 2000  & 2000 \\

\midrule
\addlinespace[0.6ex] 

\multirow{6}{*}{Sound}
& DESED           & Seg-F1 & 1153 & -    \\
& ESC-50          & Acc    & 400  & 400  \\
& FSD18-Kaggle    & mAP    & 1600 & -    \\
& FSD50k          & mAP    & 2000 & -    \\
& UrbanSound 8k   & Acc    & 873  & 873  \\
& Vocal Imitation & Acc    & 1867 & 1867 \\

\bottomrule
\end{tabular*}
\vspace{-6.0mm}
} 
\end{table}

\begin{table*}[t]
\captionsetup{font={footnotesize,stretch=0.85}}
\caption{Per-task normalized results for MLP and kNN on 20 tasks. Scores are normalized to $[0,1]$ using task-specific bounds so higher is better and reported as percentages (×100). CLAP-U replaces the Whisper encoder with CLAP-HASAT under the same training recipe. Whisper-U and Whisper-U-3 use the original Whisper Large-v3 decoder and train with one pass and three-pass replay over the same data, respectively. Highest score is in \textbf{bold}.}
\label{tab:full_results}
\vspace{-3mm}
\centering
\scriptsize
\setlength{\tabcolsep}{3.0pt}
\renewcommand{\arraystretch}{1.1}

\resizebox{\textwidth}{!}{%
\begin{tabular}{
@{\extracolsep{\fill}}
ll
*{6}{c@{\hspace{3pt}}c}
@{\hskip 8pt}
*{3}{c@{\hspace{3pt}}c}
>{\centering\arraybackslash}c@{\hspace{3pt}}>{\centering\arraybackslash}c
@{}
}
\toprule
\multirow{2}{*}{\footnotesize Domain} & \multirow{2}{*}{\footnotesize Task} &
\multicolumn{2}{c}{\makebox[5em][c]{\footnotesize wav2vec 2.0}} &
\multicolumn{2}{c}{\makebox[5em][c]{\footnotesize HuBERT}} &
\multicolumn{2}{c}{\makebox[4em][c]{\footnotesize WavLM}} &
\multicolumn{2}{c}{\makebox[4em][c]{\footnotesize BEATs}} &
\multicolumn{2}{c}{\makebox[2em][c]{\footnotesize CLAP}} &
\multicolumn{2}{c}{\makebox[3em][c]{\footnotesize Whisper}} &
\multicolumn{2}{c}{\makebox[4em][c]{\footnotesize \textit{CLAP-U}}} &
\multicolumn{2}{c}{\makebox[5em][c]{\footnotesize \textit{Whisper-U}}} &
\multicolumn{2}{c}{\makebox[5em][c]{\footnotesize \textit{Whisper-U-3}}} &
\multicolumn{2}{c}{\makebox[5em][c]{\footnotesize \textbf{Ours}}} \\
\cmidrule(lr){3-4}\cmidrule(lr){5-6}\cmidrule(lr){7-8}\cmidrule(lr){9-10}\cmidrule(lr){11-12}\cmidrule(lr){13-14}\cmidrule(lr){15-16}\cmidrule(lr){17-18}\cmidrule(lr){19-20}\cmidrule(lr){21-22}
& &
{\scriptsize MLP} & {\scriptsize kNN} &
{\scriptsize MLP} & {\scriptsize kNN} &
{\scriptsize MLP} & {\scriptsize kNN} &
{\scriptsize MLP} & {\scriptsize kNN} &
{\scriptsize MLP} & {\scriptsize kNN} &
{\scriptsize MLP} & {\scriptsize kNN} &
{\scriptsize MLP} & {\scriptsize kNN} &
{\scriptsize MLP} & {\scriptsize kNN} &
{\scriptsize MLP} & {\scriptsize kNN} &
{\scriptsize \textbf{MLP}} & {\scriptsize \textbf{kNN}} \\
\midrule

\multirow{11}{*}{\footnotesize Speech}
& ASV2015
& $94.0$ & $\mathbf{95.8}$
& $95.1$ & $88.5$
& $97.4$ & $94.6$
& $91.5$ & $81.0$
& $95.1$ & $71.2$
& $97.5$ & $90.3$
& $92.3$ & $82.9$
& $98.3$ & $95.0$
& $94.7$ & $90.6$
& $\mathbf{99.0}$ & $92.9$ \\
& CREMA-D
& $55.5$ & $21.8$
& $64.5$ & $34.0$
& $71.3$ & $28.0$
& $66.5$ & $39.7$
& $36.9$ & $25.4$
& $57.5$ & $38.2$
& $54.3$ & $49.0$
& $64.3$ & $17.8$
& $66.3$ & $30.0$
& $\mathbf{84.4}$ & $\mathbf{67.9}$ \\
& FSC
& $47.2$ & $1.6$
& $96.5$ & $3.6$
& $97.5$ & $4.3$
& $47.3$ & $5.5$
& $3.6$ & $1.6$
& $97.8$ & $23.4$
& $17.7$ & $5.4$
& $97.2$ & $1.0$
& $\mathbf{98.0}$ & $\mathbf{53.1}$
& $82.7$ & $14.6$ \\
& LibriCount
& $58.0$ & $20.8$
& $58.2$ & $18.2$
& $65.1$ & $38.8$
& $\mathbf{68.2}$ & $37.8$
& $38.8$ & $28.1$
& $59.4$ & $45.2$
& $60.9$ & $\mathbf{54.0}$
& $62.8$ & $9.1$
& $53.7$ & $41.4$
& $64.4$ & $47.2$ \\
& LS-100h
& $16.0$ & --
& $83.4$ & --
& $64.1$ & --
& $15.8$ & --
& $0.2$ & --
& $62.5$ & --
& $1.5$ & --
& $79.0$ & --
& $89.9$ & --
& $\mathbf{91.6}$ & -- \\
& LS-MF
& $95.3$ & $70.3$
& $97.8$ & $79.9$
& $97.9$ & $77.2$
& $97.5$ & $92.3$
& $90.2$ & $86.8$
& $90.8$ & $58.0$
& $97.1$ & $\mathbf{96.4}$
& $97.2$ & $53.9$
& $82.7$ & $55.9$
& $\mathbf{98.1}$ & $91.3$ \\
& RAVDESS
& $44.1$ & $15.8$
& $58.1$ & $32.9$
& $68.0$ & $25.2$
& $66.6$ & $37.9$
& $23.7$ & $21.9$
& $61.3$ & $38.5$
& $55.9$ & $45.5$
& $59.4$ & $23.9$
& $65.6$ & $36.8$
& $\mathbf{88.2}$ & $\mathbf{70.0}$ \\
& GSC
& $71.2$ & $23.2$
& $95.7$ & $28.9$
& $96.3$ & $58.3$
& $88.0$ & $49.6$
& $25.1$ & $13.7$
& $\mathbf{97.0}$ & $\mathbf{74.8}$
& $67.0$ & $49.1$
& $96.7$ & $9.8$
& $94.9$ & $73.0$
& $95.5$ & $65.4$ \\
& VocalSound
& $77.3$ & $24.1$
& $85.5$ & $35.7$
& $89.3$ & $31.1$
& $91.4$ & $75.2$
& $42.6$ & $25.9$
& $89.3$ & $57.1$
& $88.5$ & $85.8$
& $90.7$ & $39.0$
& $90.6$ & $43.3$
& $\mathbf{93.0}$ & $\mathbf{90.9}$ \\
& VoxCeleb1
& $34.0$ & $0.2$
& $58.2$ & $6.5$
& $\mathbf{65.4}$ & $8.9$
& $41.3$ & $\mathbf{13.2}$
& $4.3$ & $0.5$
& $22.0$ & $4.1$
& $13.6$ & $7.8$
& $40.9$ & $0.9$
& $33.8$ & $4.1$
& $45.5$ & $4.5$ \\
& VoxLingua33
& $55.8$ & $0.6$
& $75.0$ & $6.3$
& $86.2$ & $29.0$
& $41.9$ & $14.8$
& $7.7$ & $3.5$
& $\mathbf{97.6}$ & $\mathbf{95.8}$
& $18.8$ & $13.4$
& $97.6$ & $53.6$
& $94.0$ & $62.8$
& $89.5$ & $71.6$ \\
\midrule

\multirow{3}{*}{\footnotesize Music}
& FMA
& $47.8$ & $21.1$
& $50.5$ & $22.6$
& $52.6$ & $31.0$
& $66.0$ & $61.0$
& $30.0$ & $25.7$
& $57.4$ & $51.4$
& $66.0$ & $62.2$
& $61.0$ & $40.8$
& $61.2$ & $46.5$
& $\mathbf{68.9}$ & $\mathbf{67.3}$ \\
& GTZAN
& $63.7$ & $31.1$
& $69.6$ & $20.7$
& $74.0$ & $43.6$
& $89.8$ & $85.2$
& $40.4$ & $34.9$
& $71.2$ & $53.3$
& $84.2$ & $79.5$
& $70.5$ & $11.2$
& $82.3$ & $64.1$
& $\mathbf{94.5}$ & $\mathbf{89.5}$ \\
& NSynth
& $31.6$ & $25.1$
& $37.7$ & $35.5$
& $40.9$ & $25.0$
& $64.8$ & $61.6$
& $24.5$ & $20.1$
& $46.0$ & $12.4$
& $\mathbf{77.3}$ & $\mathbf{78.0}$
& $47.0$ & $15.0$
& $49.7$ & $38.5$
& $70.7$ & $70.0$ \\
\midrule

\multirow{6}{*}{\footnotesize Sound}
& DESED
& $31.7$ & --
& $33.5$ & --
& $38.8$ & --
& $4.2$ & --
& $2.4$ & --
& $29.4$ & --
& $20.2$ & $\mathbf{0.0}$ 
& $31.0$ & --
& $44.3$ & --
& $\mathbf{57.0}$ & -- \\
& ESC-50
& $51.9$ & $7.9$
& $56.9$ & $13.2$
& $65.7$ & $19.2$
& $95.3$ & $85.5$
& $16.3$ & $18.2$
& $62.4$ & $29.5$
& $\mathbf{97.0}$ & $\mathbf{95.6}$
& $67.0$ & $3.5$
& $80.6$ & $46.4$
& $95.8$ & $93.7$ \\
& FSD18-Kaggle
& $23.1$ & --
& $34.9$ & --
& $36.4$ & --
& $79.1$ & --
& $6.4$ & --
& $36.0$ & --
& $87.6$ & $\mathbf{0.0}$ 
& $21.8$ & --
& $55.8$ & --
& $\mathbf{90.5}$ & -- \\
& FSD50k
& $16.5$ & --
& $21.9$ & --
& $27.4$ & --
& $56.7$ & --
& $4.5$ & --
& $32.1$ & --
& $59.1$ & $\mathbf{0.0}$ 
& $34.6$ & --
& $50.3$ & --
& $\mathbf{60.0}$ & -- \\
& UrbanSound 8k
& $66.7$ & $34.8$
& $65.8$ & $34.3$
& $69.1$ & $30.6$
& $\mathbf{89.3}$ & $81.4$
& $36.4$ & $33.8$
& $71.9$ & $45.5$
& $85.4$ & $\mathbf{83.1}$
& $73.4$ & $11.5$
& $76.4$ & $54.4$
& $84.4$ & $78.3$ \\
& Vocal Imitation
& $14.5$ & $0.9$
& $17.4$ & $2.4$
& $24.5$ & $4.3$
& $24.1$ & $13.4$
& $2.5$ & $1.7$
& $24.1$ & $5.2$
& $17.2$ & $11.4$
& $26.8$ & $1.6$
& $\mathbf{28.5}$ & $10.9$
& $25.7$ & $\mathbf{14.7}$ \\
\midrule

\multicolumn{2}{l}{\scriptsize Weighted Avg.}
& $42.8$ & $27.7$
& $69.2$ & $32.6$
& $67.2$ & $37.1$
& $52.4$ & $49.2$
& $22.0$ & $27.6$
& $64.2$ & $45.7$
& $43.9$ & $53.1$
& $70.3$ & $27.8$
& $74.4$ & $47.0$
& $\mathbf{80.9}$ & $\mathbf{60.4}$ \\
\bottomrule
\end{tabular}
}

\vspace{-2.0mm}
\end{table*}

\begin{table}[t]
\centering
\captionsetup{font={footnotesize,stretch=0.85}}
\caption[Bootstrap]{Reported deviations correspond to the maximum absolute difference between the mean and the endpoints of the 95\% percentile CI, estimated using hierarchical bootstrap resampling over seeds and evaluation examples with replacement.}
\vspace{-3mm}
\label{tab:bootstrap}

\footnotesize
\setlength{\tabcolsep}{2.0pt}
\renewcommand{\arraystretch}{1}

\resizebox{\columnwidth}{!}{%
\begin{tabular}{
@{\extracolsep{\fill}}
l
*{5}{c@{\hspace{1.5pt}}c}
@{}
}
\toprule
\multirow{2}{*}{\footnotesize Task} & \multicolumn{2}{c}{{\footnotesize Whisper}} & \multicolumn{2}{c}{{\footnotesize UniWhisper}} & \multicolumn{2}{c}{{\footnotesize CLAP-Uni}} & \multicolumn{2}{c}{{\footnotesize Whisper-Uni-1}} & \multicolumn{2}{c}{{\footnotesize Whisper-Uni-3}} \\
 & {\scriptsize MLP} & {\scriptsize kNN} & {\scriptsize MLP} & {\scriptsize kNN} & {\scriptsize MLP} & {\scriptsize kNN} & {\scriptsize MLP} & {\scriptsize kNN} & {\scriptsize MLP} & {\scriptsize kNN} \\
\midrule
ASV2015 & $\scriptstyle\pm$1.7 & $\scriptstyle\pm$1.4 & $\scriptstyle\pm$2.0 & $\scriptstyle\pm$1.8 & $\scriptstyle\pm$1.1 & $\scriptstyle\pm$1.3 & $\scriptstyle\pm$2.5 & $\scriptstyle\pm$2.4 & $\scriptstyle\pm$1.4 & $\scriptstyle\pm$1.3 \\
CREMA-D & $\scriptstyle\pm$0.9 & $\scriptstyle\pm$0.8 & $\scriptstyle\pm$1.4 & $\scriptstyle\pm$1.2 & $\scriptstyle\pm$0.9 & $\scriptstyle\pm$1.1 & $\scriptstyle\pm$1.7 & $\scriptstyle\pm$0.3 & $\scriptstyle\pm$1.3 & $\scriptstyle\pm$0.4 \\
FSC & $\scriptstyle\pm$1.8 & $\scriptstyle\pm$0.4 & $\scriptstyle\pm$1.6 & $\scriptstyle\pm$0.2 & $\scriptstyle\pm$0.4 & $\scriptstyle\pm$0.1 & $\scriptstyle\pm$1.8 & $\scriptstyle\pm$0.3 & $\scriptstyle\pm$2.5 & $\scriptstyle\pm$0.7 \\
LibriCount & $\scriptstyle\pm$0.9 & $\scriptstyle\pm$0.8 & $\scriptstyle\pm$1.0 & $\scriptstyle\pm$0.9 & $\scriptstyle\pm$1.5 & $\scriptstyle\pm$0.6 & $\scriptstyle\pm$1.6 & $\scriptstyle\pm$0.2 & $\scriptstyle\pm$1.2 & $\scriptstyle\pm$0.8 \\
LS-100h & $\scriptstyle\pm$1.0 & -- & $\scriptstyle\pm$1.8 & -- & $\scriptstyle\pm$0.3 & -- & $\scriptstyle\pm$1.9 & -- & $\scriptstyle\pm$1.8 & -- \\
LS-MF & $\scriptstyle\pm$1.8 & $\scriptstyle\pm$1.0 & $\scriptstyle\pm$2.0 & $\scriptstyle\pm$2.1 & $\scriptstyle\pm$2.4 & $\scriptstyle\pm$2.5 & $\scriptstyle\pm$1.6 & $\scriptstyle\pm$1.1 & $\scriptstyle\pm$2.3 & $\scriptstyle\pm$1.0 \\
RAVDESS & $\scriptstyle\pm$1.1 & $\scriptstyle\pm$0.7 & $\scriptstyle\pm$1.9 & $\scriptstyle\pm$1.1 & $\scriptstyle\pm$1.0 & $\scriptstyle\pm$0.7 & $\scriptstyle\pm$1.3 & $\scriptstyle\pm$0.4 & $\scriptstyle\pm$1.0 & $\scriptstyle\pm$1.0 \\
GSC & $\scriptstyle\pm$1.4 & $\scriptstyle\pm$1.8 & $\scriptstyle\pm$2.5 & $\scriptstyle\pm$0.9 & $\scriptstyle\pm$1.1 & $\scriptstyle\pm$1.2 & $\scriptstyle\pm$2.1 & $\scriptstyle\pm$0.2 & $\scriptstyle\pm$2.3 & $\scriptstyle\pm$1.2 \\
VocalSound & $\scriptstyle\pm$1.8 & $\scriptstyle\pm$0.8 & $\scriptstyle\pm$1.8 & $\scriptstyle\pm$1.5 & $\scriptstyle\pm$1.4 & $\scriptstyle\pm$1.2 & $\scriptstyle\pm$1.9 & $\scriptstyle\pm$0.7 & $\scriptstyle\pm$1.9 & $\scriptstyle\pm$0.7 \\
VoxCeleb1 & $\scriptstyle\pm$0.4 & $\scriptstyle\pm$0.1 & $\scriptstyle\pm$1.1 & $\scriptstyle\pm$0.1 & $\scriptstyle\pm$0.2 & $\scriptstyle\pm$0.2 & $\scriptstyle\pm$0.8 & $\scriptstyle\pm$0.5 & $\scriptstyle\pm$0.8 & $\scriptstyle\pm$0.1 \\
VoxL33 & $\scriptstyle\pm$1.5 & $\scriptstyle\pm$1.6 & $\scriptstyle\pm$1.7 & $\scriptstyle\pm$0.9 & $\scriptstyle\pm$0.3 & $\scriptstyle\pm$0.2 & $\scriptstyle\pm$1.4 & $\scriptstyle\pm$1.2 & $\scriptstyle\pm$2.5 & $\scriptstyle\pm$1.2 \\
FMA & $\scriptstyle\pm$1.1 & $\scriptstyle\pm$0.8 & $\scriptstyle\pm$1.2 & $\scriptstyle\pm$1.3 & $\scriptstyle\pm$1.2 & $\scriptstyle\pm$1.3 & $\scriptstyle\pm$1.3 & $\scriptstyle\pm$0.7 & $\scriptstyle\pm$1.2 & $\scriptstyle\pm$1.0 \\
GTZAN & $\scriptstyle\pm$1.3 & $\scriptstyle\pm$0.9 & $\scriptstyle\pm$1.8 & $\scriptstyle\pm$2.3 & $\scriptstyle\pm$1.4 & $\scriptstyle\pm$1.5 & $\scriptstyle\pm$1.1 & $\scriptstyle\pm$0.2 & $\scriptstyle\pm$1.2 & $\scriptstyle\pm$1.2 \\
NSynth & $\scriptstyle\pm$1.3 & $\scriptstyle\pm$0.2 & $\scriptstyle\pm$1.2 & $\scriptstyle\pm$1.4 & $\scriptstyle\pm$1.7 & $\scriptstyle\pm$1.5 & $\scriptstyle\pm$0.5 & $\scriptstyle\pm$0.2 & $\scriptstyle\pm$1.2 & $\scriptstyle\pm$0.6 \\
DESED & $\scriptstyle\pm$0.6 & -- & $\scriptstyle\pm$1.3 & -- & $\scriptstyle\pm$0.2 & -- & $\scriptstyle\pm$0.5 & -- & $\scriptstyle\pm$1.0 & -- \\
ESC-50 & $\scriptstyle\pm$1.3 & $\scriptstyle\pm$0.5 & $\scriptstyle\pm$1.8 & $\scriptstyle\pm$2.1 & $\scriptstyle\pm$2.0 & $\scriptstyle\pm$2.4 & $\scriptstyle\pm$1.6 & $\scriptstyle\pm$0.1 & $\scriptstyle\pm$1.9 & $\scriptstyle\pm$1.0 \\
FSD18-K & $\scriptstyle\pm$0.7 & -- & $\scriptstyle\pm$1.2 & -- & $\scriptstyle\pm$1.0 & -- & $\scriptstyle\pm$0.5 & -- & $\scriptstyle\pm$1.4 & -- \\
FSD50k & $\scriptstyle\pm$0.7 & -- & $\scriptstyle\pm$1.1 & -- & $\scriptstyle\pm$1.3 & -- & $\scriptstyle\pm$0.7 & -- & $\scriptstyle\pm$1.2 & -- \\
UB 8k & $\scriptstyle\pm$1.4 & $\scriptstyle\pm$0.9 & $\scriptstyle\pm$1.4 & $\scriptstyle\pm$1.2 & $\scriptstyle\pm$2.1 & $\scriptstyle\pm$1.4 & $\scriptstyle\pm$1.4 & $\scriptstyle\pm$0.2 & $\scriptstyle\pm$2.2 & $\scriptstyle\pm$1.1 \\
Vocal Im & $\scriptstyle\pm$0.3 & $\scriptstyle\pm$0.1 & $\scriptstyle\pm$0.5 & $\scriptstyle\pm$0.3 & $\scriptstyle\pm$0.3 & $\scriptstyle\pm$0.2 & $\scriptstyle\pm$0.5 & $\scriptstyle\pm$0.5 & $\scriptstyle\pm$0.8 & $\scriptstyle\pm$0.2 \\
Avg. & $\scriptstyle\pm$0.5 & $\scriptstyle\pm$0.3 & $\scriptstyle\pm$0.6 & $\scriptstyle\pm$0.5 & $\scriptstyle\pm$0.2 & $\scriptstyle\pm$0.4 & $\scriptstyle\pm$0.7 & $\scriptstyle\pm$0.3 & $\scriptstyle\pm$0.8 & $\scriptstyle\pm$0.2 \\
\bottomrule
\end{tabular}
}
\vspace{-6mm}
\end{table}

\section{Results}
\label{sec:results}

We report results under the same protocols for several widely used pretrained audio encoders, including wav2vec 2.0-Large~\cite{wav2vec2}, HuBERT-Large~\cite{hubert}, WavLM-Large~\cite{wavlm}, Whisper-Large-v3~\cite{whisper}, BEATs-iter3~\cite{beats}, and CLAP-HTSAT~\cite{clap}.
Per task results are provided in Table~\ref{tab:full_results} and confidence intervals in Table~\ref{tab:bootstrap}.

\subsection{MLP results}
\label{sec:mlp_results}

Table~\ref{tab:full_results} shows that UniWhisper achieves weighted averages of 0.81 with MLP and 0.61 with kNN. 
This gain suggests that continual multi-task training with unified instruction supervision increases the amount of linearly accessible information in the representation.
We observe the largest improvements on non-speech tasks that rely on global semantic cues, such as audio tagging and captioning. Besides, UniWhisper maintains strong performance on speech-oriented tasks, including speaker and paralinguistic classification.
Compared with domain specialized baselines, UniWhisper narrows the gap on environmental sound and music tasks without requiring multi encoder feature fusion.

We also note that speech only encoders such as wav2vec and WavLM remain strong on speech heavy subsets, but they are less consistent on music and complex sound scene tasks under our unified evaluation.
Conversely, CLAP and BEATs provide strong performance on tasks aligned with their pretraining objectives, but they tend to underperform on fine grained speech tasks that depend on phonetic detail.
UniWhisper improves the cross domain balance by building on Whisper acoustics and adding supervision signals that explicitly target non-speech semantics.

\subsection{kNN results}
\label{sec:knn_results}
\vspace{-1mm}

kNN evaluation in Table~\ref{tab:full_results} largely mirrors the MLP trends while emphasizing embedding-space geometry without learning a task-specific head. UniWhisper improves kNN performance over Whisper, suggesting that instruction-style continual training yields a better organized representation space in addition to higher probe accuracy. Across baselines, we find that models optimized for global alignment such as CLAP can perform well on retrieval oriented tasks but remains weaker on fine-grained speech transfer, consistent with a retrieval-optimized encoder whose representations favor global semantic alignment over dense temporal detail. Bootstrap intervals in table~\ref{tab:bootstrap} for weighted averages are narrow, supporting the stability of these trends.

\subsection{Ablation}
\label{sec:ablation}
\vspace{-1mm}

We ablate the encoder choice and decoder design under the same instruction-style continual training pipeline.

\textbf{Encoder choice:} Inspired by CLIP~\cite{clip}, we test a CLIP-style contrastive encoder by replacing our encoder with CLAP-HASAT~\cite{laionclap} while keeping the rest of the pipeline unchanged. For fairness, CLAP-Uni bypasses CLAP's final clip-level mean pooling and uses penultimate-layer dense features: each 10\,s segment yields $h\in\mathbb{R}^{64\times 2048}$ (stride $\approx 156$\,ms). We also remove the extra temporal convolution used in UniWhisper.

As shown in Table~\ref{tab:full_results}, our instruction-style continual training substantially improves CLAP: CLAP-Uni nearly doubles the weighted average and is competitive with UniWhisper on CLAP-aligned semantic audio tasks. However, its gains are less pronounced on temporally precise speech understanding. This suggests that native temporal granularity of the backbone remains a key factor even after continual training.

\textbf{Decoder design:} We compare our pretrained LM decoder with the original Whisper decoder under the same pipeline. As shown in Table~\ref{tab:full_results}, the Whisper decoder yields slower gains and weaker semantic alignment: Whisper-Uni-1 reaches $0.70$ in MLP but drops to $0.28$ in kNN, below the original Whisper kNN score $0.46$, indicating a less well-structured embedding space despite higher probe accuracy. Multi-pass replay improves results, but at much higher cost and still below UniWhisper. This supports using a pretrained LM decoder, which better aligns instruction-style semantics and preserves representation structure under a fixed compute budget.

\section{Conclusion}
\label{sec:conclusion}

We presented UniWhisper, an efficient continual multi-task training framework for universal audio representation trained with a unified instruction and answer format for continual multi-task supervision. Starting from Whisper Large v3, we replace the original decoder with a compact pretrained language model connected through a lightweight projection adapter, enabling efficient next-token training on 38k hours of public audio. On 20 tasks spanning speech, environmental sound, and music, UniWhisper achieves weighted averages of 0.81 (MLP) and 0.61 (kNN), outperforming strong pretrained baselines while maintaining competitive speech capability.


\clearpage   
\nocite{*}   

\begingroup  
\small  
\raggedbottom

\section{Generative AI Use Disclosure}
We used a generative AI tool to assist with language editing and polishing of the manuscript, including improving grammar, clarity, and readability.
The tool was not used to generate scientific content, experimental results, or conclusions.
All coauthors reviewed the final manuscript and take full responsibility for it.
\vspace{-2pt}

\bibliographystyle{IEEEtran}
\bibliography{mybib}
\endgroup   

\end{document}